\documentclass[conference]{IEEEtran}
\usepackage{blindtext, graphicx}
\usepackage{amssymb}
\usepackage{bm}
\usepackage{tikz}
\usepackage{amsmath}

\usepackage{mathabx}
\usetikzlibrary{chains}
\usetikzlibrary{shapes}
\usetikzlibrary{shapes.misc}
\usetikzlibrary{calc}
\usetikzlibrary{arrows}
\usetikzlibrary{patterns}
\usetikzlibrary{fadings}
\usetikzlibrary{automata}
\usetikzlibrary{spy}
\usepackage{tikzscale}
\usepackage{multirow}
\usepackage{float}
\usepackage{ellipsis}
\usetikzlibrary{calc}
\usetikzlibrary{decorations.pathreplacing,decorations.markings,shapes.geometric}
\tikzset{naming/.style={align=center,font=\small}}
\tikzset{antenna/.style={insert path={-- coordinate (ant#1) ++(0,0.25) -- +(135:0.25) + (0,0) -- +(45:0.25)}}}
\tikzset{station/.style={naming,draw,shape=dart,shape border rotate=90, minimum width=10mm, minimum height=10mm,outer sep=0pt,inner sep=3pt}}
\tikzset{mobile/.style={naming,draw,shape=rectangle,minimum width=12mm,minimum height=6mm, outer sep=0pt,inner sep=3pt}}
\tikzset{radiation/.style={{decorate,decoration={expanding waves,angle=90,segment length=4pt}}}}

\usepackage{pgfplots}
\usepgfplotslibrary{fillbetween}
\usetikzlibrary{plotmarks}
\usepackage{grffile}
\newlength\figureheight
\newlength\figurewidth
\def \deffigureheight {4cm}
\def \deffigurewidth {.8\linewidth}
\usepackage{booktabs}
\usepackage[colorinlistoftodos,prependcaption]{todonotes}
\let\oldtodo\todo
\renewcommand{\todo}[1]{\oldtodo[inline]{#1}}

\makeatletter
\newcommand{\specialcell}[1]{\ifmeasuring@#1\else\omit$\displaystyle#1$\ignorespaces\fi}
\makeatother

\graphicspath{{./images/}}
\ifCLASSINFOpdf
\else
\fi

\begin{document}
% paper title
\title{Statistical Multiplexing of Computations in C-RAN with Tradeoffs in Latency and Energy}

% author names and affiliations
\author{
  \IEEEauthorblockN{Anders~E.~Kal\o r\IEEEauthorrefmark{1},
    Mauricio~I.~Agurto\IEEEauthorrefmark{1},
    Nuno~K.~Pratas\IEEEauthorrefmark{2},
    Jimmy~J.~Nielsen\IEEEauthorrefmark{2},
    Petar~Popovski\IEEEauthorrefmark{2}
  }
  Department of Electronic Systems, Aalborg University, Denmark
  \IEEEauthorblockA{\IEEEauthorrefmark{1}\{akalar12,magurt15\}@student.aau.dk, \IEEEauthorrefmark{2}\{nup,jjn,petarp\}@es.aau.dk }
}

% make the title area
\maketitle

\begin{abstract}

In the Cloud Radio Access Network (C-RAN) architecture, the baseband signals from multiple remote radio heads are processed in a centralized baseband unit (BBU) pool.
This architecture allows network operators to adapt the BBU's computational resources to the aggregate
access load experienced at the BBU, which can change in every air-interface access frame.
The degree of savings that can be achieved by adapting the resources is a tradeoff between savings, adaptation frequency, and increased queuing time.
If the time scale for adaptation of the resource multiplexing is greater than the access frame duration, then this may result in additional access latency and limit the energy savings. 
In this paper we investigate the tradeoff by considering two extreme time-scales for the resource multiplexing: (i) \emph{long-term}, where the computational resources are adapted over periods much larger than the access frame durations; (ii) \emph{short-term}, where the adaption is below the access frame duration.
We develop a general C-RAN queuing model that describes the access latency and show, for Poisson arrivals, that long-term multiplexing achieves savings comparable to short-term multiplexing, while offering low implementation complexity.

\end{abstract}

\section{Introduction}
\label{sec:introduction}

\begin{figure}
     \centering
     \setlength{\figureheight}{4cm}
     \setlength{\figurewidth}{\linewidth}
     \input{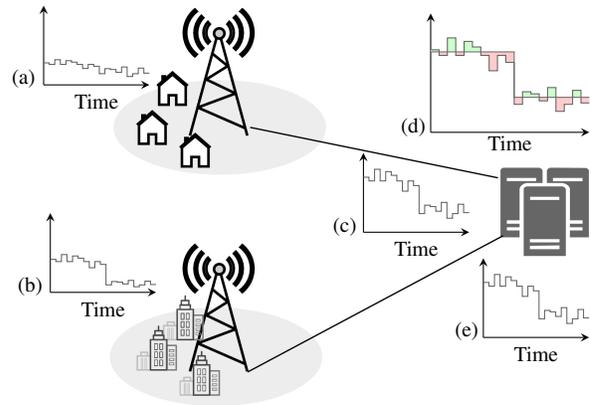}
     \setlength{\figureheight}{\deffigureheight}
     \setlength{\figurewidth}{\deffigurewidth}
     \caption{A C-RAN deployment with 2 RRHs and the application of
       BBU resource multiplexing to short-term (in frames) and
       long-term (in hours) load fluctuations. (a) and (b) Load at the
       individual RRHs. (c) Aggregate load at the BBU. (d) Long-term multiplexing.
       (e) Short-term multiplexing.}
 \label{fig:c_ran}
\end{figure}

In the Cloud Radio Access Network (C-RAN) architecture, the Remote Radio Heads (RRHs) are connected through
low latency and high capacity front-haul links to a central pool of virtual Base Band Units (BBUs), as illustrated in Fig.~\ref{fig:c_ran}.
This architecture enables the baseband signals from spatially distributed RRHs, to be partially or fully processed in the BBUs~\cite{mobile2011c}, allowing for a high level of synchronization and coordination between the RRHs. This ultimately enables spectral efficiency enhancements brought by cooperative techniques, such as coordinated multipoint (CoMP)~\cite{cloud_RAN_for_mobile_networks,comp_challenges}.

The traditional cellular network architecture, denoted as Distributed RAN (D-RAN), has the functionalities of the RRH and BBU concentrated at the base stations.
This is neither resource nor cost efficient, as the processing resources at these base stations are dimensioned to handle peak access loads.
Furthermore, the only way to reduce the energy and operating expenditures is to turn off partially or completely the base station during periods of low access load.
In a C-RAN architecture, the number of processing resources at the BBU can be chosen to take advantage of the load fluctuations across the RRHs, with the goal of reducing energy and operating costs.
Access load fluctuations are both slow and fast.
Taking a Poisson arrival perspective, \emph{slow fluctuations} refer to the average arrival rate changing over the course of the day, i.e. $\lambda(t)$ which changes in the minutes to hours scale---the so-called \emph{tidal effect}~\cite{mobile2011c}.
The \emph{fast fluctuations} refer to the instantaneous realization of
the arrival process, i.e. in the milliseconds to seconds scale.
Thus, long-term multiplexing refers to adapting the resources to $\lambda(t)$, while the short-term multiplexing refers to the capability to adapt to fast fluctuations.
In a C-RAN setting, the adaptation to slow fluctuations is achieved by enabling/disabling RRHs and the associated processing resources according to the current needs.
While this also exists in traditional radio access networks, the multiplexing gain becomes more
significant when multiple RRHs share the same computational resources, as in the BBU pool. Ideally, fast fluctuations can be taken advantage of by occupying or freeing up computational resources for the BBU pool at high frequency in an elastic cloud environment. 

Since long-term multiplexing adapts to the slow fluctuations in the
load, there may be periods where the load is higher than what can be
served by the allocated resources, as shown in the red area in the curve in Fig.~\ref{fig:c_ran}(d). These periods introduce queuing in the system and hence higher latency for the users. Similarly, there are periods where the load is lower than what the system can serve (the green area). In this region and depending on the amount of queued arrivals, some resources may be unused, thereby creating a potential for savings which the long-term multiplexing cannot achieve.
These fluctuations can be exploited by the short-term multiplexing which operates at a much higher frequency than long-term multiplexing, see Fig.~\ref{fig:c_ran}(e). In this case, the number of servers follows the
load in every frame. Since the number of active resources is limited only by the air-interface, no additional queuing latency is introduced. However, this requires quick adaptation of resources to the fluctuations, which may be very difficult in practice. 
Nevertheless, it serves as reference when measuring the potential computational resource savings.
\begin{figure}
  \centering
  \includegraphics[width=.7\linewidth]{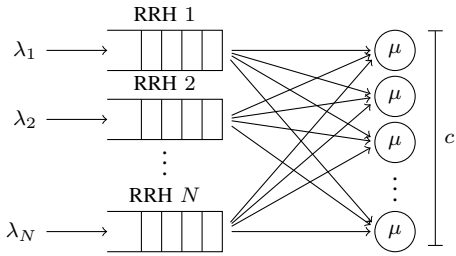}
  \caption{The transfers from each RRH are queued before they are handled by one of the $c$ shared servers in the BBU pool.}
\label{fig:sysmodel}
\end{figure}

In this paper, we analyze and evaluate the system latency introduced by the multiplexing of computational resources in the BBU pool in a frame-based admission setting, where the amount of resources is determined by the current load.
We consider two time-scales for the resource multiplexing: (i)
long-term multiplexing, where the computing resources are adapted to
the access load in an interval much higher than the air-interface
frame duration; (ii) short-term multiplexing, where the number of
computing resources can adapt faster than the frame duration. We
define a general queuing system model to study resource multiplexing
in terms of the tradeoff between latency and resource/energy
savings. We also identify the regions where the multiplexing gains at different scales are identical in terms of the achieved savings.

Latency in C-RAN has been previously
investigated~\cite{qinghan,q_analysis_bs_processing}, but only few have studied the latency incurred by computational resource
multiplexing in the BBU pool and the effect of frame-based
admission. Other studies \cite{DBLP:journals/corr/LiuZGNX14,werthmann2013multiplexing} consider the probability of missed deadlines in the case of BBU sharing, where users are served without queuing.
A scheduling framework for long-term load multiplexing is proposed in~\cite{bhaumik2012cloudiq}, where both energy savings and the probability of missing a deadline are characterized.

The remainder of the paper is organized as follows.
Section II defines the system model and in Section III we analyze the dynamics of the system latency.
Numerical results are presented and discussed in Section IV and finally the paper is concluded in Section V.

\section{System Model}\label{sec:sysmodel}

We consider $N$ RRHs connected to a shared BBU pool via a front-haul link, as presented in Fig. \ref{fig:c_ran}.
Since we are interested in the latency caused by resource multiplexing, we assume that the front-haul link is dimensioned to meet the latency constraints required by the system.
The air-interface at each RRH follows a time and frequency based frame
structure, such as in LTE~\cite{madueno2016}, of duration $F$.
The maximum number of supported concurrent user transactions in an air-interface frame is denoted by $L$.
The BBU pool shared between the $N$ RRHs consists of $c$ servers, each
able to handle one user transaction at a time (Fig. \ref{fig:sysmodel}).
Under this scheme, at most $L$ servers can serve transfers from the same RRH concurrently.
Since no more than $L\cdot N$ transfers can be active at the same time, we only consider the cases where $c \le L\cdot N$.
Each RRH has its own queue of transfers; and the available servers in
the BBU are assigned to these queues in a round-robin fashion to
provide fairness between the RRHs.

Within the time scale of an air-interface frame, we treat $c$ as constant.
In the case of long-term resource multiplexing, $c$ will only change after a large number of frames have elapsed (large enough to assume stationary conditions).
In the short-term resource multiplexing, $c$ is adapted at the beginning of each frame according to the access load.

We assume that a user transaction is composed of several uplink and downlink exchanges; corresponding to the user connection establishment to the network, the Scheduling Request (SR), the subsequent data exchanges and release of the network connection. A user transaction is completed only after all its uplink and downlink transmissions have been completed.
We model the user transaction arrivals at the $j\text{th}$ RRH as a Poisson arrival process with intensity $\lambda_j$, where the arrivals can only enter the system at the beginning of each frame.
\begin{figure}
  \centering
  \includegraphics[width=0.7\linewidth]{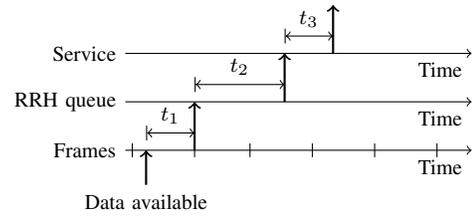}
  \caption{Transfer delays in the system.}
  \label{fig:latencies}
\end{figure}
Upon the connection establishment, the scheduling request of a user is
queued at the RRH until resources become available to initiate the
transfer and users are instantaneously informed when they are assigned
resources. 

The latency experienced by a user is illustrated in Fig.~\ref{fig:latencies}.
% These delays can be combined to obtain:
% %
% \begin{description}[\IEEEsetlabelwidth{Transfer time}\IEEEusemathlabelsep]
%   \item[\emph{Queuing time}] $t_2$,
%   \item[\emph{Waiting time}] $t_1+t_2$,
%   \item[\emph{Transfer time}] $t_1+t_2+t_3$.
% \end{description}
% %
The delay $t_1$ corresponds to the time from when data is available to be transferred at the user device until the scheduling request is transmitted in the beginning of the following frame.
$t_2$ corresponds to the time spent in the RRH queue, i.e. the time from reception of the scheduling request until the first resource is granted. 
The last delay, denoted as $t_3$, is the service time, i.e. the time it takes for the user to transmit its data.
We assume that the amount of requested resources in a scheduling request follows an exponential distribution with rate $\mu$~\cite{ftp3gpp}. In this way the amount of requested resources may exceed one frame, in which case the transfer will span multiple frames.

\subsection{Resource/Energy Savings}
\label{sub:sysmodel_savings}

In long-term multiplexing, the servers are adapted to the mean arrival rate.
We assume that some target delay $\tau$ exists with a certain
reliability $\zeta$ (e.g. less than 1 frame period queuing time with
probability 99\%) and that the minimum number of servers
required to fulfill this requirement are allocated. %(at most $L\cdot N$).

For short-term multiplexing, we assume that servers can be turned off when they are idle, and are only turned back on when needed. If servers are turned on and off instantaneously and at any time, the achievable savings would be equal to the mean idle time, $1-\rho$.
However, we consider a more realistic scenario, where servers can be turned on and off only in the beginning of each frame.
Specifically, the number of transfers in the system is observed immediately after the arrival of SRs in the beginning of each frame, and the servers required to serve them are allocated while the remaining servers are turned off.
The number of allocated servers in the BBU is given by
\begin{equation}
  c = \sum_{j=1}^N \min(L, l_j)
\end{equation}
where $l_j$ is the number of transfers in the system (ongoing and queued) from RRH $j$.
%The above allocation scheme achieves the optimal savings ($1-\rho$) only in the case where the service incurred periods end immediately before the beginning of a new frame.

\section{Analysis} % (fold)
\label{sec:analysis}

In this section we obtain the probability distributions describing the latency in the system.
Based on these distributions we find expressions for potential energy/resource savings.
We start by defining the conditions required to ensure stability in the network.
Then we characterize the long-term resource multiplexing through an approximate Markov chain model which describes the metrics of interest.
Finally, we present the short-term resource multiplexing, where the number of servers $c$ at the BBU pool is dynamically adopted in each frame to the fast fluctuations of the arriving traffic.

\subsection{Stability Conditions} % (fold)
\label{sub:stability_conditions}

The stability condition for the described system is given in terms of the utilization $\rho$.
Since the utilization is both limited by the total number of servers in the BBU, $c$, and the number of concurrent transactions, $L$, there are two conditions that must be satisfied for stability.
First, the total number of arrivals must be less than what the $c$
servers in the BBU pool can handle \eqref{eq:rho1}. Second, the arrivals at each RRH must be below what is supported by the air-interface \eqref{eq:rho2}:
\begin{align}
  &\rho_{\text{BBU}}= \frac{1}{c\mu F}\sum_{j=1}^N \lambda_{j} < 1,\label{eq:rho1}\\
  &\rho_{\text{RRH}j}= \frac{\lambda_{j}}{L\mu F} < 1,\ \  \forall\, j=1\ldots N.\label{eq:rho2}
\end{align}
When $\rho_{\text{BBU}}\ge 1$ or $\rho_{\text{RRH}j}\ge 1$, the queue grows to infinity.

% subsection stability_conditions (end)

\subsection{Long-term Resource Multiplexing} % (fold)
\label{sub:long_term_resource_multiplexing}

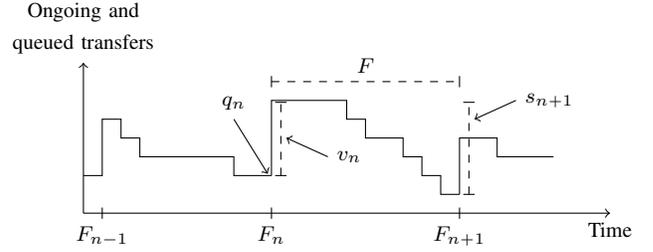
\begin{figure}
  \centering
  \begin{tikzpicture}[scale=0.25]
  
  % Coordinate system
  \draw[->] (0,0) -- (0,8);
  \node[anchor=south,align=center] at (0,8) {\footnotesize Ongoing
    and\\\footnotesize queued transfers};
  \draw[->] (0,0) -- (28,0);
  \node[anchor=north] at (28,0) {\footnotesize Time};

  % First arrival (5 jobs)
  \draw (0,2) -- (1,2) -- (1,5) -- (2,5) -- (2,4) -- (3,4) -- (3,3) -- (8,3) -- (8,2) --
  (10,2);
  \draw (1,-0.3) -- (1,0.3) node[anchor=north] at +(0,-0.5)
  {\footnotesize $F_{n-1}$};
  
  % Second arrival (4 jobs)
  \draw (10,2) -- (10,6) -- (14,6) -- (14,5) -- (15,5) -- (15,4) --
  (17,4) -- (17,3) -- (18,3) -- (18,2) -- (19,2) -- (19,1) -- (20,1);
  \draw (10,-0.3) -- (10,0.3) node[anchor=north] at +(0,-0.5)
  {\footnotesize $F_{n}$};
  \draw[<-] (9.8,2.1) -- (8,5) node[anchor=south] {\footnotesize $q_{n}$};
  \draw (10.2,2) -- (10.8,2);
  \draw[dashed] (10.5,2) -- (10.5,6);
  \draw (10.2,5.9) -- (10.8,5.9);
  \draw[<-] (10.8,4) -- (13,3) node[anchor=west] {\footnotesize $v_{n}$};

  % Third arrival (2 jobs)
  \draw (20,1) -- (20,4) -- (22,4) -- (22,3) -- (25,3);
  \draw (20,-0.3) -- (20,0.3) node[anchor=north] at +(0,-0.5)
  {\footnotesize $F_{n+1}$};
  \draw (20.2,1) -- (20.8,1);
  \draw[dashed] (20.5,1) -- (20.5,6);
  \draw (20.2,5.9) -- (20.8,5.9);
  \draw[<-] (20.8,5) -- (23,6) node[anchor=west] {\footnotesize $s_{n+1}$};

  % F
  \draw[dashed] (10,7) -- (20,7) node[midway,above] {\footnotesize $F$};
  \draw (10,6.7) -- (10,7.3);
  \draw (20,6.7) -- (20,7.3);
\end{tikzpicture}
  \caption{The notation used in the Markov chain model.}
  \label{fig:notation}
\end{figure}

In this subsection, we present a Markov chain model which approximates
the latencies introduced by queuing in the system under
stationarity.
%\footnote{The results in Section~\ref{sec:numerical_results} show that the model is a good approximation.}.
Motivated by the round-robin scheduling, we analyze the RRHs
individually with a fixed number of servers proportional to the
arrival rates.
The number of servers used for analyzing RRH $j$ is given as:
\begin{equation}\label{eq:numserversapprox}
  \hat{c}_j = \left\lfloor\frac{\lambda_{j}}{\sum_n\lambda_{n}} c\right\rfloor
\end{equation}
where $c$ is the number of servers in BBU pool and $\lfloor x\rfloor$ denotes the largest integer less than or equal to $x$. To simplify the notation, we shall refer to $\hat{c}_j$ simply as $c$ and $\lambda_j$ as $\lambda$ in the remainder of this section and refer to Fig.~\ref{fig:notation} for the Markov chain symbol notation.

We consider a discrete-time Markov chain that is observed in the beginning of each frame, immediately prior to the arrival of SRs, denoted as $\{F_n\}$.
The states are indexed by the number of transfers (ongoing and queued) in the system $q_n$.
Let $v_n$ denote the number of SRs arriving in frame $n$ specified by the Poisson probability mass function
\begin{equation}
  p_{v_n}(v_n)=\frac{\lambda^{v_n}e^{-\lambda}}{v_n!}, \qquad v_n\ge 0,
\end{equation}
and let $s_{n+1}$ be the number of transfers completed between $F_n$ and $F_{n+1}$ (Fig. \ref{fig:notation}). We form the basic relation 
\begin{equation*}
 q_{n+1} = q_n + v_n - s_{n+1}.
\end{equation*}

We seek the transition probabilities $p_{i,j}$ in the Markov chain such that
\begin{equation*}
  p_{i,j}=\Pr(q_{n+1}=j|q_{n}=i).
\end{equation*}
%

%Throughout the analysis, we use the memoryless property of the exponentially distributed service time, which allows us to ignore the previous states.
For convenience, we analyze the transition probabilities conditioned on the number of arriving SRs in a frame since this quantity is independent of the system state:
\begin{equation*}
  \psi_{i+k,j}=\Pr(q_{n+1}=j|q_{n}=i,v_n=k).
\end{equation*}

From $\psi_{i+k,j}$ we obtain $p_{i,j}$ by marginalizing over the number of arrivals $v_n$:
\begin{equation}\label{eq:pij}
  p_{i,j} = \begin{cases}
    \sum_{k=0}^{\infty} \psi_{i+k,j}p_{v_n}(k) & i+k\ge j
    \ge 0,\\
    0 & \text{otherwise.}
  \end{cases}
\end{equation}

We consider the following three cases of $\psi_{i+k,j}$:
\begin{enumerate}
  \item $i+k\le c$ corresponding to when all arriving SRs are immediately served.
  \item $i+k> c,\, j\ge c$ corresponding to when some of the arriving SRs are placed in the queue, and all servers remain busy during the entire frame.
  \item $i+k> c,\, j<c$  corresponding to when some of the arriving SRs are placed in the queue, but at least one server is idle in the beginning of the succeeding frame.
\end{enumerate}

When $i+k\le c$ all $k$ arriving SRs are immediately served. Exactly
$i+k-j$ transfers will complete service within the frame
period, or equivalently, exactly $j$ out of $i+k$ will \emph{not}
complete transfer. Since the service time is exponential, the
probability that a transfer will not complete service within the
frame period is $e^{-\mu F}$, and hence we obtain
\begin{equation*}\label{eq:case1}
  \psi_{i+k,j}=\binom{i+k}{j}e^{-\mu F j}(1-e^{-\mu F})^{i+k-j},
  \qquad i+k\le c.
\end{equation*}

For case 2, where $i+k>c, j\ge c$, exactly $i+k-c$ SRs
are placed in the queue and all $c$
servers will remain busy throughout the frame period during which
$i+k-j$ transfers will complete service. Since all servers are busy in the
entire frame period, the number of completed transfers is Poisson
distributed with rate $c\mu F$:
\begin{equation*}
  \psi_{i+k,j}=\frac{(c\mu F)^{i+k-j}e^{-c\mu F}}{(i+k-j)!},
  \qquad i+k> c,\, j\ge c.
\end{equation*}

When $i+k>c, j<c$, one or more servers are idle prior to the subsequent frame. As in the previous case, $i+k-c$ SRs are initially placed in the queue and all servers are busy, but after $i+k-c+1$ transfers have been completed, some servers remain idle for the remaining frame.

Let $t\le 1$ denote the time (in frame durations) until $i+k-c+1$ transfers have completed. 
$t$ is a sum of $i+k-c+1$ independent and identically distributed exponential random variables and hence is Erlang distributed~\cite{kleinrockv1} with density function
\begin{equation*}
  p(t)=\frac{(c\mu F)^{i+k-c+1}t^{i+k-c}e^{-c\mu F t}}{(i+k-c)!}.
\end{equation*}

The probability that $j<c$ transfers remain in service after $\xi$ frame periods, assuming the total number of transfers in the system is less than $c$ is given by
\begin{equation*}
  p(j|\xi)=\binom{c-1}{j}e^{-\xi\mu F j}(1-e^{-\xi\mu F})^{c-j-1}.
\end{equation*}
Marginalizing over $\xi=1-t$ we obtain the final expression for $\psi_{i+k,j}$:
\begin{align*}
  \psi_{i+k,j}=&\int_0^1
  \binom{c-1}{j}e^{-(1-t)j\mu F}(1-e^{-(1-t)\mu F})^{c-j-1}\nonumber\\
  &\frac{(c\mu F)^{i+k-c+1}t^{i+k-c}e^{-c\mu F t}}{(i+k-c)!}\,
     dt\nonumber\\
  =&\binom{c-1}{j}\frac{e^{-j\mu F}(c\mu F)^{i+k-c+1}}{(i+k-c)!}\nonumber\\
     &\int_0^1e^{-t\mu F(j-c))}(1-e^{-(1-t)\mu F})^{c-j-1}
     t^{i+k-c}\, dt,\\
     &\specialcell{\hspace{12.6em} \text{$i+k> c,\, j<c$.}}
\end{align*}

To obtain the stationary queuing time distribution we first seek the
stationary queue length distribution.
Let $\bm{P}=\begin{bmatrix}p_{i,j}\end{bmatrix}$ denote the transition
matrix and $\bm{\pi}=\begin{bmatrix}\pi_i\end{bmatrix}$ be a vector of
state probabilities. Since the Markov chain is irreducible and
aperiodic, the stationary state distribution $\bm{\pi}$ is given by
\begin{equation*}
  \bm{\pi}=\lim_{n\to\infty}\bm{\pi}^{(0)} \bm{P}^n
\end{equation*}
where $\bm{\pi}^{(0)}$ is the initial state distribution.
We obtain $\bm{\pi}$ by imposing a finite queue length $M$ and multiplying iteratively by $\bm{P}$ until convergence.
Equation \eqref{eq:pij} then becomes:
\begin{multline*}
  p_{i,j} = \sum_{k=0}^{\infty}
  p_{v_n}(k)\psi_{i+\min(k,c+M-i),j},\\
  \qquad i+k\ge j \ge 0,\, i\le c+M.
\end{multline*}

From $\bm{\pi}$ we may obtain the distribution of the queuing time $t_2$ under stationarity.
Let $q'_{n} = q_n + v_n$ be the number of transfers in the system immediately \emph{after} arrival of SRs. Since we assume stationarity, we omit the time index and write $q' = q + v$.
We may factorize the queuing time distribution as
\begin{multline*}
  p_{t_2}(t_2) =
  \sum_{q}\sum_{q'}\sum_{l}p_{t_2|l}(t_2|l)\Pr(l|q,q')\Pr(q'|q)\pi_q,\\
  t_2\ge 0,\, c+M\ge q'\ge q\ge 0,
\end{multline*}
where $p_{t_2|l}(t_2|l)$ is the density function of the queuing time conditioned on an SR arriving to state $l$ and $\Pr(l|q,q')$ is the probability of arriving to state $l$ given that the newly arrived SRs occupy states
$q+1,\ldots,q'$. Two cases of $p(t_2|l)$ exists: when the SR arrives to an idle server ($l\le c$), and when it arrives to the queue ($l>c$).
In the former case, the SR is immediately served and the queuing time
is 0. In the latter case, the queuing time is Erlang distributed with
parameters $l-c$ and $c\mu$:
\begin{equation*}
  p_{t_2|l}(t_2|l)=\begin{cases}
  \frac{(c\mu)^{l-c}t_2^{l-c-1}e^{-c\mu t_2}}{(l-c-1)!} &l>c, \\
  \delta(t_2) & \text{otherwise}
  \end{cases}
\end{equation*}
where $\delta(x)$ is the Dirac delta function.
It is equally likely for an SR to arrive to any of the states between $q+1$ and $q'$, hence $\Pr(l|q,q')$ is a discrete uniform distribution between $q+1$ and $q'$, i.e. $\Pr(l|q,q')=(q'-q)^{-1}$. $\Pr(q'|q)$ is
obtained by truncating the Poisson distribution at the queue size limit $M$, 
\begin{equation*}
  \Pr(q'|q)=\begin{cases}
    p_{v_n}(q'-q)&q'<M,\\
    1-\sum_{n=q}^{M-1} p_{v_n}(n-q)&q'=M,\\
    0&\text{otherwise.}
    \end{cases}
\end{equation*}

We may obtain the system time, $t_2+t_3$ by the convolution of the queuing time and the service time density functions. Similarly, the transfer time $t_1+t_2+t_3$ can be obtained by convolution of the densities for $t_2+t_3$ and $t_1$. Since $t_1$ is uniformly distributed in the range $[0,F]$ we obtain
\begin{align*}
  p_{t_2+t_3}(t)&=\int_0^{t} p_{t_2}(x)\mu e^{-\mu(t-x)}\, dx,\\
  p_{t_1+t_2+t_3}(t)&=\frac{1}{F}\int_{0}^{F} p_{t_2+t_3}(t-x)\, dx.
\end{align*}

The normalized savings (server-hours) in the long-term multiplexing
scheme, as defined in~\ref{sub:sysmodel_savings}, is expressed as
\begin{equation}\label{eq:lt_savings}
  S_{\text{LT}} = \frac{1}{L\cdot N}\min\{c: \Pr(t_2 < \tau)\ge \zeta \}
\end{equation}
where $L\cdot N$ is the maximum number of servers in the BBU pool and $\tau$ and $\zeta$ are design parameters.

% subsection long_term_resource_multiplexing (end)

\subsection{Short-term Resource Multiplexing} % (fold)
\label{sub:short_term_resource_multiplexing}

Recall that in the short-term multiplexing we assume instant adaptation to
the active and queued transfers in the beginning each frame
(see~\ref{sub:sysmodel_savings}). The number of transfers
in the system immediately after arrival of the scheduling requests in
the beginning of the frame is given by
\[
  \Pr(q')=\sum_{q}\Pr(q'|q)\pi_q.
\]
As we assume that we can instantly switch servers on and off, we
may obtain the expected number of active servers by
marginalizing over $q'$. By further using the fact that at most
$c$ servers can
be active at the same time, and normalizing by
$c$, the expected savings of short-term
multiplexing are,
\begin{equation}\label{eq:expectedsavings}
  \mathbb{E}[S_{\text{ST}}]=1-\frac{1}{c}
  \sum_{q'}\min(q',c)\Pr(q')
\end{equation}
One case of $c$ which is particularly interesting
is where the maximum number of servers is used in the BBU
pool and the RRHs have equal arrival rates, i.e. $c=L$.
Since we are adapting in the beginning of each frame, the frame length
has high impact on the savings. Specifically, when the frame length is
short, we can adapt more often and the servers will be inactive for
shorter time. This is also clear from \eqref{eq:expectedsavings} where
$q'$ will be lower (in a stochastic ordering sense) when the frame length is shorter due to fewer
arrivals per frame.

% subsection short_term_resource_multiplexing (end)

% section analysis (end)

%
\label{sec:results}
%\begin{figure*}[t]
%\setlength{\figureheight}{4cm}
%\setlength{\figurewidth}{.25\linewidth}
%\centering
%\definecolor{mycolor1}{rgb}{0.00000,0.44700,0.74100}%
%\definecolor{mycolor2}{rgb}{0.85000,0.32500,0.09800}%
%\definecolor{mycolor3}{rgb}{0.92900,0.69400,0.12500}%
%\definecolor{mycolor4}{rgb}{0.49400,0.18400,0.55600}%
%\definecolor{mycolor5}{rgb}{0.46600,0.67400,0.18800}%
%\definecolor{mycolor6}{rgb}{0.30100,0.74500,0.93300}%
%  \ref{named}
%  \begin{subfigure}[t]{0.3\textwidth}
%    \input{images/queueing_time.tex}
%    \caption{Mean queuing time ($t_1$).}%Waiting time vs. the number of servers for different $\lambda$.}
%    \label{fig:mean_queueing_vs_servers}
%  \end{subfigure}
%  \hfill
%  \begin{subfigure}[t]{0.3\textwidth}
%    \input{images/waiting_time.tex}
%    \caption{Mean waiting time ($t_1+t_2$).}%Waiting time vs. the number of servers for different $\lambda$.}
%    \label{fig:mean_waiting_vs_servers}
%  \end{subfigure}
%  \hfill
%  \begin{subfigure}[t]{0.3\textwidth}
%    \input{images/transfer_time.tex}
%    \caption{Mean transfer time ($t_1+t_2+t_3$).}%Waiting time vs. the number of servers for different $\lambda$.}
%    \label{fig:mean_latency_vs_servers}
%  \end{subfigure}
%  \caption{Mean latencies in different parts of the system for
%    different numbers of servers. \todo{We have to revise this plot for clarity, the legend and the number of curves is too confusing.}}
%  \label{fig:mean_time_vs_servers}
%  \setlength{\figureheight}{\deffigureheight}
%  \setlength{\figurewidth}{\deffigurewidth}
%\end{figure*}
%
%
\begin{table}[t]
  \centering
  \caption{\textsc{Parameters considered in the evaluation}}
  \label{tab:evalparams}
  \begin{tabular}{lcc}
    \toprule
    \textbf{Parameter} & \textbf{Symbol} & \textbf{Value} \\\midrule
    Frame duration & $F$ & $10$ \\
    Maximum concurrent transactions per RRH & $L$ & $25$ \\
    Number of RRHs & $N$ & $2$ \\
    Mean number of requested resources & $1/\mu$ & $5$ \\
    \bottomrule
  \end{tabular}
\end{table}

\section{Numerical Results} % (fold)
\label{sec:numerical_results}

In this section we present the numerical results of the long-term and
short-term resource multiplexing approaches.
We consider a system with the parameters specified in Table \ref{tab:evalparams}.
We study the case with two RRHs with equal arrival rates
$\lambda_1=\lambda_2$, as this is sufficient to show
the dynamics of the resource multiplexing.
%In the following we denote $\tau$ as the target queuing time and $\zeta$ as the percentile where $\tau$ is met.
%
\begin{figure}
  \centering
  % This file was created by matlab2tikz.
%
%The latest updates can be retrieved from
%  http://www.mathworks.com/matlabcentral/fileexchange/22022-matlab2tikz-matlab2tikz
%where you can also make suggestions and rate matlab2tikz.
%
\definecolor{mycolor3}{rgb}{0.00000,0.44700,0.74100}%
\definecolor{mycolor5}{rgb}{0.63500,0.07800,0.18400}%
\definecolor{mycolor2}{rgb}{0.92900,0.69400,0.12500}%
\definecolor{mycolor4}{rgb}{0.49400,0.18400,0.55600}%
\definecolor{mycolor1}{rgb}{0.46600,0.67400,0.18800}%
\begin{tikzpicture}

\begin{axis}[%
width=0.951\figurewidth,
height=\figureheight,
at={(0\figurewidth,0\figureheight)},
scale only axis,
xmin=0,
xmax=50,
xlabel={$c$},
ymin=0,
ymax=10,
font=\footnotesize,
ylabel={99 percentile queuing time},
xlabel style={yshift=5pt},
ylabel style={yshift=-15pt},
axis background/.style={fill=white},
axis x line*=bottom,
axis y line*=left,
legend style={legend cell align=left,align=left,draw=white!15!black,font=\scriptsize},
grid=both
]
\addplot [color=mycolor1,solid,mark=asterisk,mark options={solid}]
  table[row sep=crcr]{%
50	4.7743e-11\\
48	2.3718e-10\\
46	1.1387e-09\\
44	5.2767e-09\\
42	2.3577e-08\\
40	1.0145e-07\\
38	4.1982e-07\\
36	1.6687e-06\\
34	6.3616e-06\\
32	2.3228e-05\\
30	8.1109e-05\\
28	0.00027046\\
26	0.00085999\\
24	0.0026043\\
22	0.0075043\\
20	0.020567\\
18	0.053658\\
16	0.1336\\
14	0.31945\\
12	0.74344\\
10	1.7356\\
8	4.3913\\
6	16.551\\
5	20\\
};
\addlegendentry{$\lambda=5$};

\addplot [color=mycolor2,solid,mark=asterisk,mark options={solid}]
  table[row sep=crcr]{%
50	1.8963e-05\\
48	4.8635e-05\\
46	0.00012091\\
44	0.00029121\\
42	0.00067902\\
40	0.001532\\
38	0.0033428\\
36	0.0070522\\
34	0.014381\\
32	0.028352\\
30	0.054053\\
28	0.099746\\
26	0.1784\\
24	0.30996\\
22	0.52498\\
20	0.87166\\
18	1.4326\\
16	2.3747\\
14	4.1406\\
12	8.6643\\
11	20\\
};
\addlegendentry{$\lambda=10$};

\addplot [color=mycolor3,solid,mark=asterisk,mark options={solid}]
  table[row sep=crcr]{%
50	0.087291\\
48	0.12692\\
46	0.18127\\
44	0.25454\\
42	0.35181\\
40	0.47926\\
38	0.64456\\
36	0.85754\\
34	1.1315\\
32	1.486\\
30	1.9524\\
28	2.5882\\
26	3.5195\\
24	5.1014\\
22	8.9726\\
21	20\\
};
\addlegendentry{$\lambda=20$};

\addplot [color=mycolor4,solid,mark=asterisk,mark options={solid}]
  table[row sep=crcr]{%
50	0.98861\\
48	1.1957\\
46	1.4426\\
44	1.7397\\
42	2.103\\
40	2.5593\\
38	3.1596\\
36	4.0176\\
34	5.4598\\
32	9.0215\\
31	20\\
};
\addlegendentry{$\lambda=30$};

\addplot [color=mycolor5,dashed,thick]
  table[row sep=crcr]{%
0	1\\
50	1\\
};
\addlegendentry{$\tau=1$};

\addplot [color=mycolor1,dashdotted,thick]
  table[row sep=crcr]{%
5	0\\
5	100\\
};
\addplot [color=mycolor2,dashdotted,thick]
  table[row sep=crcr]{%
10	0\\
10	100\\
};
\addplot [color=mycolor3,dashdotted,thick]
  table[row sep=crcr]{%
20	0\\
20	100\\
};
\addplot [color=mycolor4,dashdotted,thick]
  table[row sep=crcr]{%
30	0\\
30	100\\
};

% \draw (axis cs:12,1) circle(1);
% \draw (axis cs:20,1) circle(1);
% \draw (axis cs:36,1) circle(1);
% \draw (axis cs:50,1) circle(1);

% \draw (axis cs:12,2) -- (axis cs:10,2.3);
% \draw (axis cs:20,2) -- (axis cs:20,2.3);
% \draw (axis cs:36,2) -- (axis cs:33,2.3);
% \draw (axis cs:50,2) -- (axis cs:48,2.3);

% \node at (axis cs:6,2) [anchor=south] {\tiny $\rho=0.42$};
% \node at (axis cs:20,2) [anchor=south] {\tiny$\rho=0.50$};
% \node at (axis cs:33,2) [anchor=south] {\tiny$\rho=0.56$};
% \node at (axis cs:46,2) [anchor=south] {\tiny$\rho=0.60$};

\end{axis}
\end{tikzpicture}%
  \caption{The 99-percentile queuing delay vs. the number of servers for
    different arrival rates, where $\lambda=\lambda_1=\lambda_2$. The dashed
    horizontal line indicates $\tau=1$, and the dash-dotted lines show the asymptotes.}
  \label{fig:queueing_vs_c}
\end{figure}
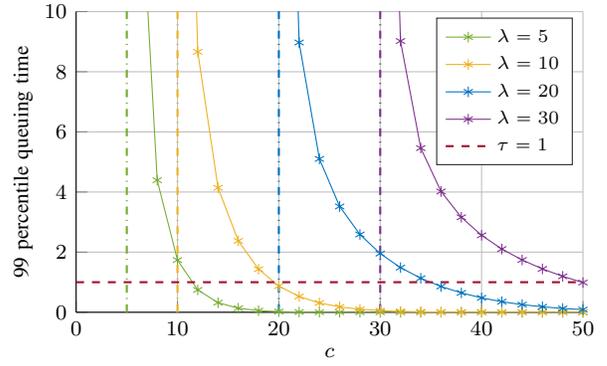

\subsection{Resource Multiplexing Savings} % (fold)
\label{sub:long_term_resource_multiplexingResults}

The plot presented in Fig. \ref{fig:queueing_vs_c} shows the 99-percentile queuing time experienced in a system with long-term multiplexing for a different number of 
available servers and arrival rates. The horizontal dashed line
indicates $\tau=1$. The intersection between the curves and $\tau=1$
corresponds to the number of servers required to achieve a 99-percentile queuing time of $\tau=1$, i.e. $1/10$ frame duration.

The lower bound of the queuing time, for all considered $\lambda$, occurs when $c=50$.
On the other hand, there is a minimum number of servers required to keep the system stable that obeys the condition in eq.~\eqref{eq:rho1}.
This point is reflected in the vertical asymptotes in the plot.
The fact that the queuing time only decreases slightly when the number of servers increases indicates that savings can be done with only limited increased latency.
This motivates the long-term multiplexing scheme where the number of servers in the BBU pool is adapted to the slowly varying mean arrival rate.
For instance, in the considered case with arrival rates $\lambda_1=10,\lambda_2=10$, only 20 servers are
required to provide 99-percentile queuing time $\tau = 1$,
which introduces considerable savings when compared to the case where all the servers are active (i.e. $c=50$). Hence, long-term multiplexing provides high savings in this case.
\begin{figure}
  \centering
  \input{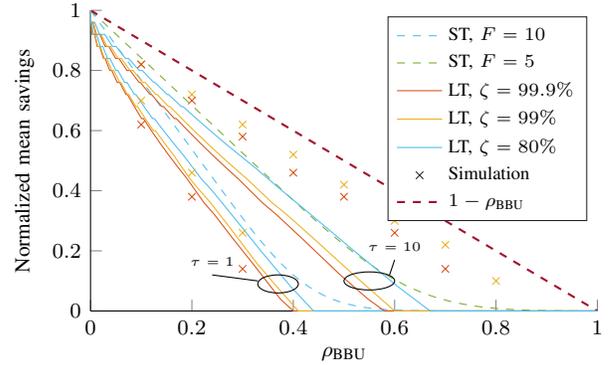}
  \caption{Long-term (LT) and short-term (ST) server-hour savings for different mean
    arrival rates (reflected in $\rho_{\text{BBU}}$) and $\zeta$-percentile
    queuing times. $1-\rho_{\text{BBU}}$ for the case with $c=50$
    provides an upper bound on the savings. The simulation results
    correspond to the 99- and 99.9-percentile cases.}
  \label{fig:longterm_savings}
\end{figure}

The savings (server-hours normalized by the server-hours in the baseline case with 50 servers)
which can be achieved using long-term and short-term multiplexing are illustrated in Fig. \ref{fig:longterm_savings}.
For long-term multiplexing we consider the minimum number of servers
required to provide a queuing delay of $\tau=1$ and $\tau=10$ at
different percentiles $\zeta$.
We obtain this number by calculating
the queuing time distribution functions for different values of $c$
and choose the minimum that satisfies the expression in
\eqref{eq:lt_savings}.

The simulation results, shown for the 99- and 99.9-percentiles, reveal that the derived analytical model fits well for $\tau=1$ but overestimates the number of servers required in the case of $\tau=10$.
This effect comes from the analysis considering each RRH queue separately, which leads to a lower statistical
multiplexing gain compared to the actual system where high queuing delays are less likely.

As shown in Fig. \ref{fig:longterm_savings}, the case with $\tau=10$ allows for higher savings since we allow the queue to be larger and hence can reduce the number of active servers.
Likewise, a low percentile allows for higher savings since we allow the queuing time to exceed $\tau$ with higher probability.
This reflects that high savings come at the cost of an increased queuing delay.
However, increasing the percentile only leads a to minor decrease in savings, which indicates that significant savings can be achieved with long-term multiplexing while maintaining a very low latency.

The long-term savings approach the upper bound provided by the normalized idle time, $1-\rho_{\text{BBU}}$, as $\tau\to\infty$.
Independently of $\zeta$ and $\tau$, a low utilization
$\rho_{\text{BBU}}$ (i.e. a low arrival rate) also leads to higher
savings since less servers are required to provide the desired queuing
delay. Even by slightly increasing the queuing time, notably $\tau=1$, it is possible to achieve significant savings when the utilization is low.

The savings achieved by short-term multiplexing are also shown in Fig. \ref{fig:longterm_savings} for frame durations of $F=10$ and $F=5$. 
Interestingly, we see that if the delay requirements are not too strict and $\tau=10$ can be accepted, then the long-term multiplexing provides higher savings than the short term multiplexing with $F=10$, which is the same frame length as in LTE.
Even if only $\tau=1$ can be accepted, the savings of long-term multiplexing are only slightly lower than the short-term multiplexing for $F=10$.
If a reduced frame length of $F$ is used, larger savings can be achieved; approaching the upper bound $1-\rho_{\text{BBU}}$ as $F\to1$.

We note that savings comparable to the short-term multiplexing can be achieved using long-term multiplexing, while still offering guarantees of low latency.
With the outlook to smaller frame sizes in 5G \cite{tullberg2014towards} it is unlikely that servers can be turned on and off fast enough to enable short-term multiplexing. Even if it became possible, short frames and faster resource adaptation causes high levels of signaling overhead and high
complexity, meaning that long-term multiplexing would anyways be
preferred. An exception is the support of ultra-reliable low latency
communications (URLLC), where latency violations are not
acceptable. However, URLLC is not well suited for round-robin scheduling, as considered in this paper, and will likely require latency sensitive schedulers.

% section numerical_results (end)

\section{Conclusion}
\label{sec:conlusion}

This paper studies the latency and energy tradeoffs in computational
multiplexing in C-RAN.
We identify two multiplexing time-scales: (i) long-term multiplexing, where the mean arrival rate varies over the time of day; and (ii) short-term multiplexing where the statistical multiplexing in the arrivals within each frame is exploited.
The long-term multiplexing introduces additional queuing delay, but has low implementation complexity.
Short-term multiplexing does not add queuing delay, but is difficult to realize in practice since it requires switching resources on and off at a very high frequency.
We propose a general system model where user transfers are modelled as jobs in a queuing system with servers shared among several RRHs.

We show that both multiplexing schemes can provide significant resource savings. Furthermore, it is possible to achieve long-term multiplexing savings that are comparable to those in the short-term while still maintaining a low queuing latency in high percentiles.
This suggests that long-term multiplexing provides a good tradeoff
between resource savings and realization complexity.
%Under this scheme there is a tradeoff between
%frame length, and hence signaling overhead, and the savings.
%We show that considerable savings can be achieved in the 
%frame lengths while maintaining minimal latency.
%This suggests that resource multiplexing
%may be achieved without increasing latency at all.

%\NP{What have we learned with this investigation?}

\section*{Acknowledgment}
This work was performed partly in the framework of H2020 project FANTASTIC-5G (ICT-671660) and partly by the European Research Council Consolidator Grant Nr. 648382.
The authors acknowledge the contributions of the colleagues in FANTASTIC-5G.

% Investigating more realistic adaptive schemes, in which
% servers do not start and stop instantaneously, could be considered for
% future work. Finding the optimal adaptive schemes in terms of latency
% remains a topic of research.

%\begin{figure}
%  \centering
%  \input{images/savings_pctile_F5.tex}
%  \caption{Server-hour savings when the frame duration is $F=5$.}
%  \label{fig:longterm_savings_F5}
%\end{figure}
% \begin{figure}
%   \centering
%   \input{images/shorttermsavings_mu.tex}
%   \caption{Short-term server-hour savings for different mean
%     arrival rates (expressed in $\rho_{\text{BBU}}$) and requested
%     resources $1/\mu$. The frame length is fixed at $F=10$. The idle time provides an
%     upper bound on the savings.}
%   \label{fig:adap_savings_mu}
% \end{figure}

% Can use something like this to put references on a page
% by themselves when using endfloat and the captionsoff option.
\ifCLASSOPTIONcaptionsoff
\newpage
\fi

% trigger a \newpage just before the given reference
% number - used to balance the columns on the last page
% adjust value as needed - may need to be readjusted if
% the document is modified later
% \IEEEtriggeratref{8}
% The "triggered" command can be changed if desired:
% \IEEEtriggercmd{\enlargethispage{-5in}}

% references section

% can use a bibliography generated by BibTeX as a .bbl file
% BibTeX documentation can be easily obtained at:
% http://www.ctan.org/tex-archive/biblio/bibtex/contrib/doc/
% The IEEEtran BibTeX style support page is at:
% http://www.michaelshell.org/tex/ieeetran/bibtex/
\bibliographystyle{IEEEtran}
% argument is your BibTeX string definitions and bibliography database(s)

% biography section
% 
% If you have an EPS/PDF photo (graphicx package needed) extra braces are
% needed around the contents of the optional argument to biography to prevent
% the LaTeX parser from getting confused when it sees the complicated
% \includegraphics command within an optional argument. (You could create
% your own custom macro containing the \includegraphics command to make things
% simpler here.)
% \begin{biography}[{\includegraphics[width=1in,height=1.25in,clip,keepaspectratio]{mshell}}]{Michael Shell}
%   or if you just want to reserve a space for a photo:

%   \begin{IEEEbiography}[{\includegraphics[width=1in,height=1.25in,clip,keepaspectratio]{picture}}]{John Doe}
%     \blindtext
%   \end{IEEEbiography}

%   You can push biographies down or up by placing
%   a \vfill before or after them. The appropriate
%   use of \vfill depends on what kind of text is
%   on the last page and whether or not the columns
%   are being equalized.

%   \vfill

%   Can be used to pull up biographies so that the bottom of the last one
%   is flush with the other column.
%   \enlargethispage{-5in}

%   that's all folks
\end{document}